# Instabilities and bifurcations in turbulent porous media flow

Vishal Srikanth[1] and Andrey V. Kuznetsov[1]

**Abstract**

Microscale turbulent flow in porous media is conducive to the development of flow instabilities due to strong vortical and shearing flow occurring within the pore space. When the flow instabilities around individual solid obstacles interact with numerous others within the porous medium, unique symmetry-breaking phenomena emerge as a result. This paper focuses on investigations of the vortex dynamics and flow instabilities behind solid obstacles in porous media, emphasizing how solid obstacle geometry and porosity influence both microscale and macroscale flow behavior. Two distinct symmetry-breaking mechanisms were identified in different porosity ranges. In low porosity media (< 0.8), a "deviatory flow" phenomenon occurs, where the macroscale flow deviates from the direction of applied pressure gradient at Reynolds numbers above 500. Deviatory flow is a source of macroscale Reynolds stress anisotropy, which is counterbalanced by a diminished vortex core size. In the intermediate porosity regime (0.8-0.95), a "jetting flow" mechanism creates asymmetric microscale velocity channels in the pore space through temporally biased vortex shedding, occurring during the transition to turbulence. Both symmetry-breaking phenomena are critically influenced by solid obstacle shape, porosity, and Reynolds number. Circularity of solid obstacle geometry and an adequately high Reynolds number provide critical conditions for symmetry-breaking, whereas porosity can be used to parametrize the degree of symmetry-breaking. This paper provides fundamental insights into the intricate flow dynamics in porous media, offering a comprehensive understanding of how microscale vortex interactions generate macroscale flow asymmetries across different geometric configurations.

**Article Highlights**

- Turbulent flow in porous media exhibits dual characteristics of both classical internal and external flows.
- Persistent microscale vortex production and transport induces secondary instabilities and flow bifurcations.
- Unique flow instabilities and symmetry-breaking phenomena occur at low and intermediate values of porosity.

1. Introduction

Turbulent flow in porous media represents a distinct class of fluid dynamics problems with significant relevance across numerous engineering and natural systems. From pebble-bed nuclear reactors and heat exchangers to crude oil extraction and forest fire modeling, the complex interaction between fluid flow and porous media creates unique flow phenomena that cannot be adequately captured by classical turbulence

E-mail address: avkuznet@ncsu.edu (A.V. Kuznetsov)
[1] Department of Mechanical and Aerospace Engineering, North Carolina State University, Raleigh, NC 27695, USA

theories. Turbulence in porous media is characterized by its dual nature, exhibiting features of both internal and external turbulent flows, with the solid obstacles and pore geometry imposing fundamental constraints on the spatiotemporal scales of turbulent structures. Turbulence emerges at very low Reynolds numbers in porous media due to the continuous production of vortical structures behind the numerous solid obstacles that make up the porous medium, making it prone to flow instability. Transition to turbulence was experimentally characterized by Seguin et al. (1998) and observed at pore scale Reynolds numbers of 190 for packed bed porous media. The Reynolds number was calculated using the physical diameter of the pores and the double averaged (over space and time) fluid velocity.

The field of turbulence modeling in porous media has evolved considerably over the past few decades. Early approaches employed Reynolds Averaging (RA) combined with Volume Average Theory (VAT) to develop macroscale turbulence models. This RA-VAT methodology, as detailed by Lage et al. (2007), de Lemos (2012), and Vafai (2015), provided a framework for understanding bulk flow behavior while making simplifying assumptions about the underlying microscale physics. Initial investigations of microscale flow relied heavily on Reynolds Averaged Navier-Stokes (RANS) simulations, but these were limited by inherent modeling errors as noted by Iacovides et al. (2014). Advancements in computational capabilities have enabled more sophisticated approaches such as Large Eddy Simulation (LES) and Direct Numerical Simulation (DNS) to analyze the microscale flow physics. Heavy emphasis was placed on the statistical description of the turbulent flow by averaging microscale behavior to the macroscale and developing macroscale closure models. These studies have made significant progress in understanding the relevant scales of microscale turbulent flow in porous media (He et al. 2019; He et al. 2018; Jin et al. 2015; Jin and Kuznetsov 2017; Rao and Jin 2022; Uth et al. 2016), as well as the RA flow distribution and budget of turbulence kinetic energy in porous media (Chu et al. 2018; Kuwata and Suga 2017; Kuwata and Suga 2015; Kuwata and Suga 2013).

However, few studies focus on the microscale flow in porous media from the perspective of temporal variation of the flow, vortex dynamics, and vortex interactions. Microscale vortices play a crucial role in turbulence transport in porous media with coherent turbulent structures being formed as a direct evolution of these vortical structures. Microscale vortex dynamics is strongly influenced by solid obstacle geometry, specifically porosity ($\varphi$) playing a crucial role in determining the nature of flow instabilities in porous media. Three distinct flow regimes can be identified based on porosity: low porosity ($\varphi < 0.8$), where pore spaces resemble interconnected channels with vortices often limited to the recirculating type; intermediate porosity ($0.8 < \varphi < 0.95$), where sufficient space exists for shedding vortices and the von Karman instability to develop while still interacting with downstream obstacles; and high porosity ($\varphi > 0.95$), where flow patterns around individual obstacles become largely independent of neighboring solid obstacles.

The microscale vortices generated behind solid obstacles are the source of the largest turbulent structures in porous media. Unlike classical flow around isolated cylinders, the von Karman vortex streets in porous media are constrained by the confining geometry, leading to unique patterns of instability. These instabilities can manifest as symmetry-breaking phenomena, where the flow field deviates from expected symmetric patterns despite symmetric geometry and boundary conditions. Additionally, bifurcations such as the Hopf bifurcation have been observed in both laminar and turbulent regimes, leading to oscillatory behavior in drag forces and increased tortuosity of flow streamlines.

Understanding the mechanisms of these instabilities and bifurcations is not merely of theoretical interest but has profound implications for accurate modeling of transport processes in porous media. The flow patterns determined by these instabilities directly influence important macroscale parameters such as macroscale drag forces, Reynolds stress distributions, and heat transfer rates. As industrial and environmental applications continue to demand more accurate predictive models, investigating the fundamental physics of turbulent instabilities in porous media becomes increasingly essential.

## 2. Flow instabilities in single and tandem solid obstacles

### 2.1. Single solid obstacles

Many porous media that are relevant in the turbulent flow regime are formed by the arrangement of individual solid obstacles, such as heat exchanger fins and tube banks, trees in canopy flows, and tall buildings in urban settlements. Consider the classical case of von Karman vortex instability formed behind a single cylindrical solid obstacle. The phenomenon of von Karman vortex shedding for the flow around a single cylinder was first presented in von Karman (1911). The von Karman vortices are formed by a pair of shear layers surrounding the solid obstacle surface interacting in the recirculation region of the solid obstacle. These transient vortices are then advected downstream by the flow creating an alternating vortex street in the wake of the solid obstacle. Schaefer and Eskinazi (1959) identified three basic regions formed in the vortex street, namely the formation, stable, and unstable regions. Boghosian and Cassel (2016) present a vortex-shedding mechanism to describe two dimensional von Karman vortices in the stable region based on two coincident conditions: (1) the existence of a location with zero momentum, and (2) the presence of a net force having a positive divergence. The vortex street then terminates in the unstable region leading to turbulent flow behavior. Following the unstable region, vortex street breakdown and the formation of a secondary vortex street have also been studied extensively (Durgin and Karlsson 1971; Dynnikova et al. 2016; Kumar and Mittal 2012). However, this secondary mechanism is not of significant interest in the context of porous media since the pore volume is confined preventing the formation of fully developed vortex streets.

Transition from steady to unsteady flow around circular cylinders occurs at a Reynolds number ($Re$) of 46, which results in formation of the 2D vortex street (Williamson 1989). The wake undergoes a further transition at Re ~ 190 to three-dimensional vortex structures even before the flow transitions to turbulence. A chaotic wake that is a characteristic of turbulence is observed at Re > 260 (Williamson 1996). In confined geometry, such as the presence of nearby solid walls, vortex formation can experience several bifurcating modes ranging from steady and symmetric to unsteady and asymmetric (Boghosian and Cassel 2010). Most relevant to the flow in a porous medium is the phenomenon of symmetry-breaking of the vortex street behind a single solid obstacle caused by the proximity of the solid wall (Thompson et al. 2021). In laminar flows, when the gap between the circular cylinder and the solid wall becomes less than the diameter of the cylinder, the vortices are shed such that they are self-propelled away from the solid wall (Rao et al. 2013). As the separation distance decreases, the small separation gap introduces blockage that decreases circulation resulting in asymmetric vortices. The asymmetric vortex wake persists in the near-wall flow in the turbulent flow regime, as well as for other solid obstacle shapes such as spheres and cuboids (Edegbe et al. 2024; Li et al. 2025; Thompson et al. 2021; Wang and Tan 2008). Vortex shedding is also known to interact with boundary layers in near-wall flow and lead to early laminar-turbulence transition at the location where the vortex wake meets the boundary layer (Kyriakides et al. 2012; Kyriakides et al. 1995).

## 2.2. Tandem solid obstacles

When multiple solid obstacles are arranged together, secondary flow instabilities and asymmetric flow emerge both due to proximal solid surfaces as well as the interaction of the vortex wakes surrounding the individual solid obstacles. Studies on in-line two-tandem circular cylinders revealed that the dynamics of vortex shedding behind the downstream cylinder of the two-tandem cylinders is characterized by two frequencies (Alam and Zhou 2008). Interaction of the vortex wake behind the upstream cylinder with the downstream cylinder led to a decrease in the characteristic vortex shedding frequency behind the downstream solid obstacle. This phenomenon is influenced by the separation distance between the cylinders, which modulates the interaction of free shear layers formed behind the cylinder with the neighboring cylinder. Three regimes are identified based on the separation distance between the cylinders in Zhou and Yiu (2006): overshoot, reattachment, and co-shedding regimes, depending on whether the shear layers surrounding the upstream cylinder overshoot, reattach downstream, or form vortices. Increasing the separation distance decreased the Strouhal number of the vortex shedding behind the downstream cylinder ultimately leading to diminished vortex structure interaction in the co-shedding regime. Carmo et al. (2010) performed numerical stability analysis to describe how secondary instabilities evolve behind the tandem cylinders and demonstrated how interacting vortex wakes behind tandem cylinders are fundamentally different than isolated vortex wakes behind single cylinders. Flow around staggered tandem cylinders experience 4 regimes depending on both separation distance between the cylinders in both radial and azimuthal directions (Zhou et al. 2009). Dual frequency vortex shedding was observed for intermediate separation distances (1.2-2.2 times the cylinder diameter) and angles of orientation (10º-75º) with different Strouhal numbers for the individual cylinders. Deflection of the vortex wake from the plane of symmetry of the individual solid obstacles is observed in these cases. At large separation distances (>2.2 times the cylinder diameter), dual frequency dynamics is observed at low Reynolds numbers but distinct dual frequencies vanish at higher Reynolds numbers due to an extending shear layer that reattaches on the downstream cylinder.

Dual frequency dynamics is also observed within the 4 regimes of flow observed for square cylinders placed side by side in the transverse direction (Alam et al. 2011; Alam and Zhou 2013; Ma et al. 2017). When the cylinders were placed side by side with a separation distance 1.3-2.2 times the cylinder diameter, wide and narrow streets are formed which divides the wake creating low and high frequencies, respectively. Further separation between the solid obstacles at 2.2-3 times the cylinder diameter leads to a transition regime, where a three-vortex system is formed behind the two cylinders. In this regime, the phase alignment of the vortex shedding behind the individual solid obstacles creates a channeling effect two vortices are transported together. Phase alignment of the vortex shedding disappears at higher separation distances of 4-6 times the cylinder diameter.

When the number of circular cylindrical obstacles in-line in a row is increased, the asymmetric instantaneous flow distributions and large amplitude oscillations of the wake in the downstream solid obstacles are observed (Liang et al. 2009). At critical values of solid obstacle spacing (3.6 times cylinder diameter), the flow exhibits asymmetric vortex shedding characterized by 180º phase difference between neighboring solid obstacles, which disappears when the wake behind the solid obstacles become virtually independent of one another. Asymmetric vortex shedding mechanisms arising from the vortex wake interaction between neighboring solid obstacles increases the RMS fluctuations of the drag force creating greater instability in the flow when compared to independent vortex shedding mechanisms. Hosseini et al. (2020) reported that when significantly large number of circular cylinders form an array, a two-row structure vortex shedding occurs due to convectively unstable flow forming above and below the row of solid obstacles. Significant magnitude of fluctuating lift force is observed after the 20$^{th}$ cylinder in the row. Asymmetric vortex shedding in the instantaneous flow is also observed for square cylinder solid obstacles (Bao et al. 2012). Here, the transition from symmetric to asymmetric wake is attributed to vortex impingement and a 'jet' flow mechanism that streams fluid from one side of the solid obstacle gap to the other.

### 3. Flow instabilities in porous media

Tube banks are a very commonly encountered form of porous media where individual tubes are arranged in in-line, staggered, or random arrangements. These types of porous media can be found in engineered systems like heat exchangers or in naturally occurring systems like flow inside canopies. Interesting fluid dynamics emerges when large (or infinite) arrays of solid obstacles are encountered by the flow caused by the interacting vortex systems and shear layers inside the void space. Ziada and Oengören (2000) experimentally investigated flow periodicities in both in-line and staggered solid obstacle arrangements and noted that three Strouhal numbers characterized temporal flow periodicity in staggered solid obstacle arrangements. Unique Strouhal numbers were associated with shear layer and vortex shedding instabilities at low Reynolds number that give rise to a single Strouhal number at high Reynolds number. The Strouhal number was also dependent on the space in between the solid obstacles. For in-line tube banks, instabilities arise primarily from jetting flow occurring in between the rows of solid obstacles (Ziada 2006). Phase difference in the velocity fluctuations across the transverse plane and asymmetric velocity distributions are observed in these cases. Flow instabilities also occur in porous media composed of in-line square cylinder solid obstacles in the form of a Hopf bifurcation (Agnaou et al. 2016; Zhang 2008). In this phenomenon, flow in between the solid obstacles begins to periodically oscillate over time above a critical value of Reynolds number. These unsteady oscillations are evident from the lift and drag forces acting on the solid obstacle recorded over time, as well as the instantaneous velocity distribution. These studies also reported dual frequency dynamics of the flow that terminates with a greater degree of disorder (more frequencies) appearing as the Reynolds number increases. It is interesting to note that for both single, tandem, and tube bank arrays of solid obstacles, the secondary flow instabilities emerge at laminar flow Reynolds numbers and persist in the turbulent flow regime until it is eventually overshadowed by highly energetic and chaotic vortex street dynamics at high Reynolds numbers. The secondary instabilities and symmetry-breaking flow phenomena discussed so far primarily influence the magnitudes of the macroscale drag force and velocity, as well as the microscale velocity and pressure. However, these instabilities and symmetry-breaking bifurcations disappear after Reynolds averaging resulting in symmetric and periodic Reynolds averaged microscale flow distributions.

There are a few instances reported in turbulent flow in porous media where the symmetry-breaking bifurcation extends to the Reynolds averaged flow leading to pitchfork type of bifurcations where multiple solution states are simultaneously possible for the same flow conditions. Yang & Wang (2000) numerically studied a bifurcating microscale flow behavior that occurs in periodic porous media. Symmetric or asymmetric vortex rotation modes are observed in the microscale flow streamlines beyond a critical value of inlet velocity affecting the magnitude of macroscale velocity. Aiba et al. (1982) experimentally studied the pressure distributions on the surface of the solid obstacle in in-line tube banks with varied tube spacings. They discovered that the pressure distribution was asymmetrical for low solid obstacle spacing and this asymmetry persisted after Reynolds averaging. The phenomenon was further investigated numerically by Iacovides *et al.* (2013, 2014) and West *et al.* (2014) in the context of cross flow in in-line circular tube bank heat exchangers. The asymmetry was said to increase with decreasing tube spacing and was attributed to the flow's tendency to follow the path of least resistance. The phenomenon was also examined by Abed & Afgan (2017) who suggested asymmetry in the location of the "separation shear layer" and the blockage influence of downstream tubes as possible causes. Two instances of flow instabilities resulting symmetry-breaking phenomena at low and intermediate porosities were studied in detail with a focus on analyzing the underlying microscale flow physics (Srikanth et al. 2021; Srikanth and Kuznetsov 2024b; Srikanth and Kuznetsov 2024a). The studies demonstrated how the symmetry-breaking impacts macroscale variables like drag force and the macroscale Reynolds stress tensor. The remaining discussion will focus on the key results from these studies.

### 3.1. Symmetry-breaking in the low porosity flow regime ($\varphi<0.8$)

Porous media with low porosity experience a unique flow behavior where the solid obstacles produce recirculating vortical structures similar to the flow behind a single solid obstacle. However, the void space between the solid obstacles is small leading to flow characteristics typically found in classical internal flows such as channel flows. In this low porosity flow regime, a symmetry-breaking phenomenon occurs in porous media where the macroscale flow deviates from the direction of applied pressure gradient (Abed and Afgan 2017; Iacovides et al. 2014; Srikanth et al. 2021). The phenomenon occurs during turbulent flow ($Re > 500$) in porous media composed of circular obstacles in the porosity range 0.43-0.72. The symmetry breakdown originated through the amplification of flow instabilities during vortex shedding and represents a pitchfork bifurcation with multiple possible modes. Key sensitivity factors for the occurrence of this phenomenon included porosity, obstacle shape, and Reynolds number. As a result of flow deviation sustaining in the Reynolds averaged flow, macroscale turbulence anisotropy is observed where the macroscale Reynolds stress tensor's principal axis aligns with neither geometric symmetry axes nor flow direction. This type of symmetry-breaking is referred to as "deviatory flow" hereafter.

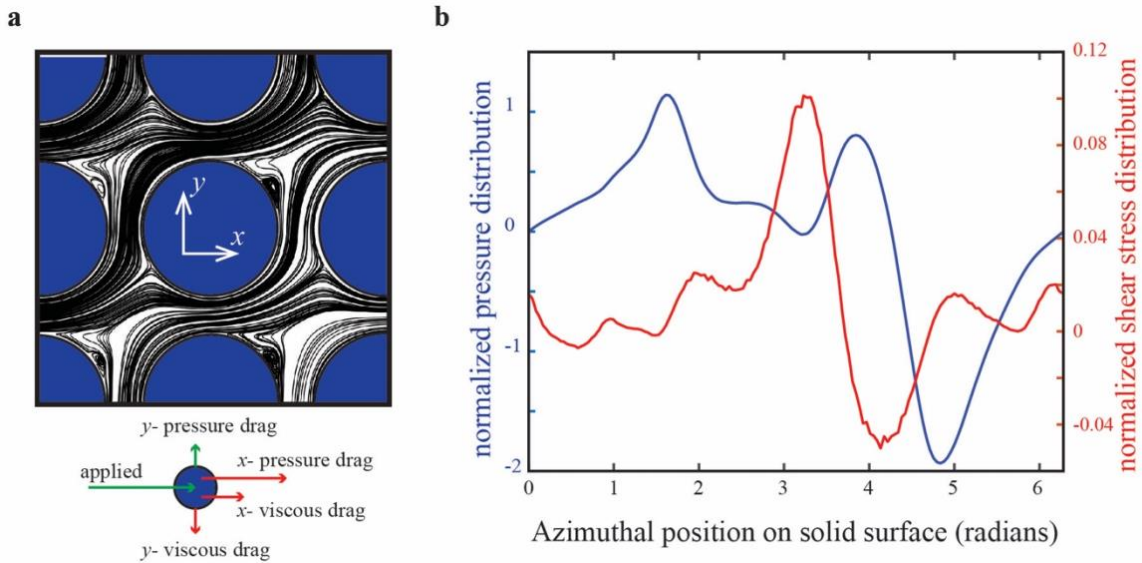

**Fig. 1.** (a) Reynolds averaged flow streamlines in periodic porous media showing the deviatory flow caused by symmetry-breaking at low porosity (Re = 1000, circular cylinder solid obstacles, $\varphi = 0.5$). The Reynolds averaged viscous and pressure drag components are not aligned as a result to asymmetric stress distributions (b).

Analyzing the microscale flow for $\varphi = 0.5$ (circular cylinder solid obstacles), dual frequency dynamics and three-dimensional vortical flow emerged at $Re = 225$, which was followed by the formation of hairpin vortex structures at $Re = 300$. At these Reynolds numbers, even though secondary flow instabilities emerged, the Reynolds averaged flow distribution was symmetric. When the Reynolds number is increased from 300 to 500, the flow transitions from a symmetric to an asymmetric state leading to the emergence of deviatory flow. Deviatory flow is illustrated in Fig. 1 along with solid obstacle surface force vectors. The resultant macroscale deviatory flow formed an angle $\theta_{macro}$ with the direction of applied pressure gradient. Remarkably, despite the asymmetric stress distribution on the solid obstacle surface, the Reynolds-averaged deviatory flow maintained momentum conservation in the direction perpendicular to the applied pressure gradient (the transverse direction). This was because the viscous and pressure components of transverse drag force had equal magnitudes but acted in opposing directions producing a net zero force. By studying the transient flow development from symmetric to asymmetric flow, it was found that the symmetry-breaking arises from low-frequency instabilities created by competing macroscale inertial and pressure forces. This instability generates high-amplitude pressure force oscillations, simultaneously overlaid with high-frequency oscillations from the von Karman vortex shedding instability.

The onset of symmetry-breaking is characterized by the disappearance of discrete frequency oscillations from the power spectrum, replaced by a continuous band of turbulence-characteristic frequencies (Fig. 2). When the Reynolds number increased beyond the critical threshold for symmetry-breaking, pressure force oscillation amplitudes reach a critical point, inducing localized transverse flow at the microscale level. Post-symmetry-breaking, these oscillation amplitudes decrease, indicating that symmetry-breaking stabilizes the highly unsteady flow instabilities present at subcritical Reynolds numbers. Even though transition to turbulence preceded symmetry-breaking and is not directly tied to the origin of the phenomenon, turbulent flow enables symmetry-breakdown by causing micro-vortex breakdown and increasing the adverse pressure gradient magnitude localized in the converging portion of the pore geometry. These two conditions induced the symmetry-breaking process from the microscale level, which develops into a macroscale phenomenon over time. The phase differences in pressure forces acting on individual obstacles create different deviatory flow modes due to chaotic turbulent motions. In infinitely periodic porous media, countless combinations of symmetry-breaking directions behind each obstacle are possible, each producing a unique deviatory flow mode influenced by the vortex shedding phase behind each obstacle and its immediate neighbors. These modes only present in the direction of applied pressure gradient with nearly identical flow patterns observed across the transverse direction.

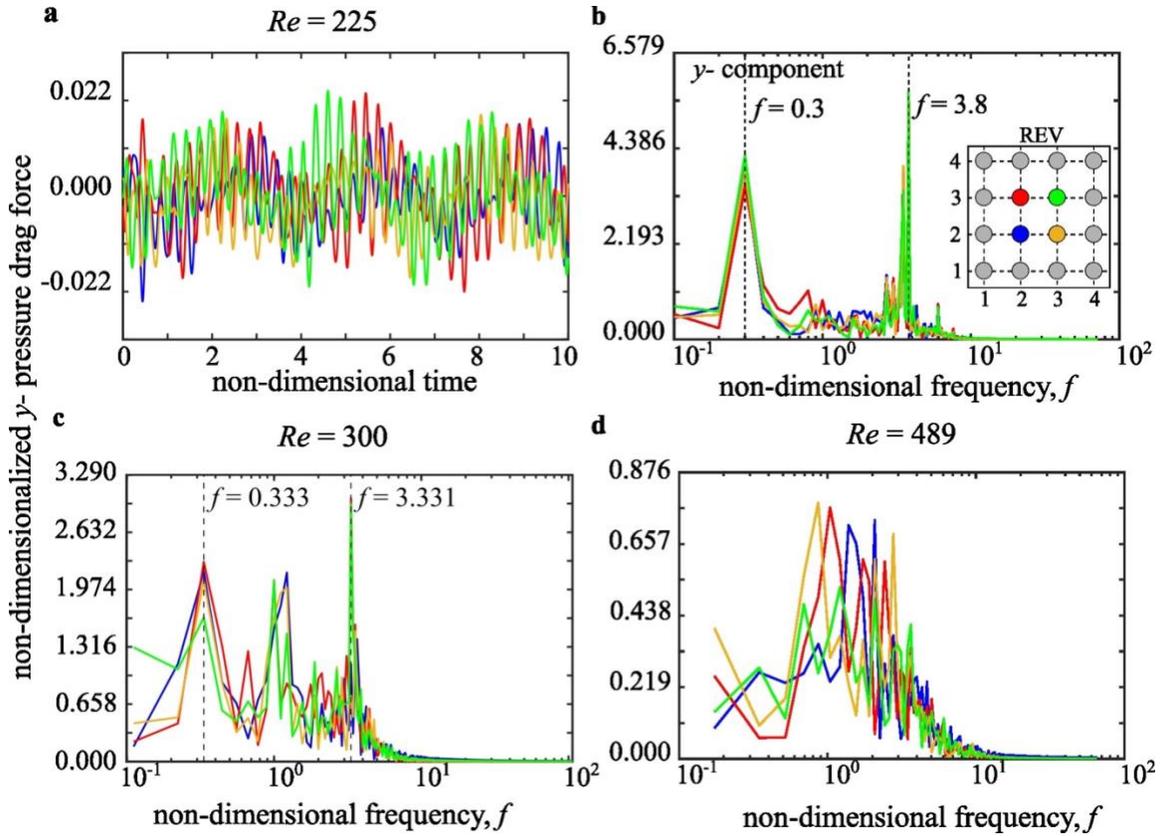

**Fig. 2.** Transverse instantaneous pressure drag force (*y*- direction) acting on the individual solid obstacle surfaces before and after symmetry breaking show the disappearance of dual frequency dynamics when deviatory flow emerges.

The deviatory flow from symmetry-breaking transforms macroscale turbulence properties in porous media. Microscale flow field asymmetry produces non-zero, non-diagonal components in the macroscale Reynolds Stress Tensor. The principal axes of this tensor form a three-dimensional angle with the geometric axes, aligning with neither the macroscale velocity vector nor the geometric axes—signifying complete flow symmetry breakdown (Fig. 3(a)). Despite the symmetry-breaking of the macroscale Reynolds Stress Tensor, the non-diagonal Reynolds Stress terms are orders of magnitude smaller than the diagonal terms. Thus, when the turbulence anisotropy is visualized using an invariant map (Emory and Iaccarino 2014), three dimensional turbulence approaching isotropic behavior is observed when the porosity was decreased (Fig. 3(b)) in spite of deviatory flow. When deviatory flow occurs, the separation point of the flow is moved downstream at lower porosities resulting in the formation of smaller micro-vortices (Fig. 3(c)). These small micro-vortices broke down into more uniformly shaped coherent turbulent structures when compared to the tubular and hairpin shaped coherent structures that exist at higher porosities (Srikanth et al. 2021). The orientation of the principal axes of the macroscale Reynolds Stress Tensor indicates the possibility of macroscale turbulence anisotropy, but the change in the vortical structures at the microscale level due to symmetry-breaking promotes more isotropic behavior by equalizing the diagonal terms of the macroscale Reynolds Stress Tensor.

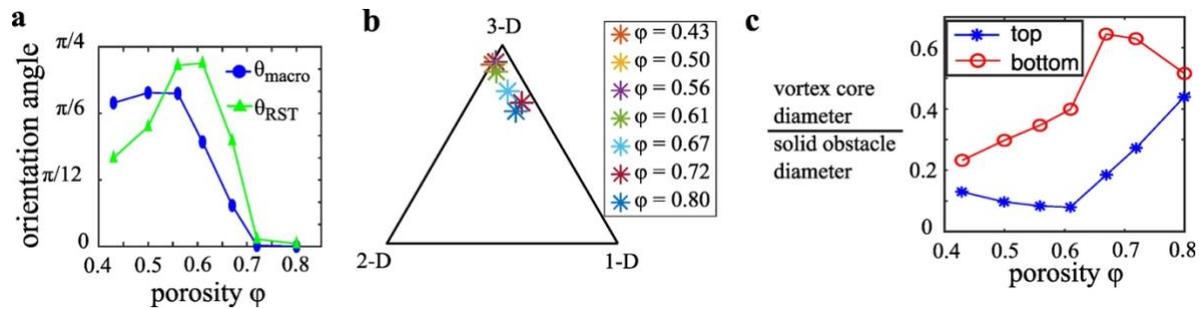

**Fig. 3.** (a) Orientations of both the macroscale flow direction ($\theta_{macro}$) and the macroscale Reynolds Stress Tensor ($\theta_{RST}$) with respect to the axes of geometric axes are non-zero. (b) Anisotropy invariant map visualization of turbulence anisotropy at different porosities. (c) Vortex core diameter variation with porosity (normalized with respect to obstacle diameter).

Considering the interesting mechanism of deviatory flow and its impact on microscale turbulent structures and macroscale turbulent transport, the occurrence of symmetry-breaking under conditions of different porosity, Reynolds number, and obstacle shape were analyzed. Critical values of porosity and Reynolds number served as necessary conditions for the occurrence of deviatory flow since the initial symmetry breaking at the microscale level is driven by high-magnitude lateral favorable pressure gradients in confined geometries. When the porosity of the porous medium is decreased below the critical value, two configurations of microscale flow separation about the geometric symmetry plane exist, each with unique properties due to differences in flow separation and stagnation locations (Fig. 4(a)). Symmetry-breaking was unaltered by Reynolds number above the critical value where the phenomenon persisted for Reynolds numbers from 500-10000 and possibly beyond. Solid obstacle shape played a crucial role in the occurrence of symmetry-breaking where circularity of the solid obstacle surface determined whether flow separation points can shift to allow deviatory flow to occur. For this reason, triangular, square, and diamond shaped solid obstacles did not produce the necessary conditions for deviatory flow (Fig. 4(b)), but polygonal (n>5), spherical, and rough circular cylinders experienced deviatory flow (Fig. 4(a)). Deviatory flow was also observed for both in-line and staggered solid obstacle arrangements. Note that for diamond shaped solid obstacles, microscale symmetry-breaking of the vortices observed by Yang and Wang (2000) is present.

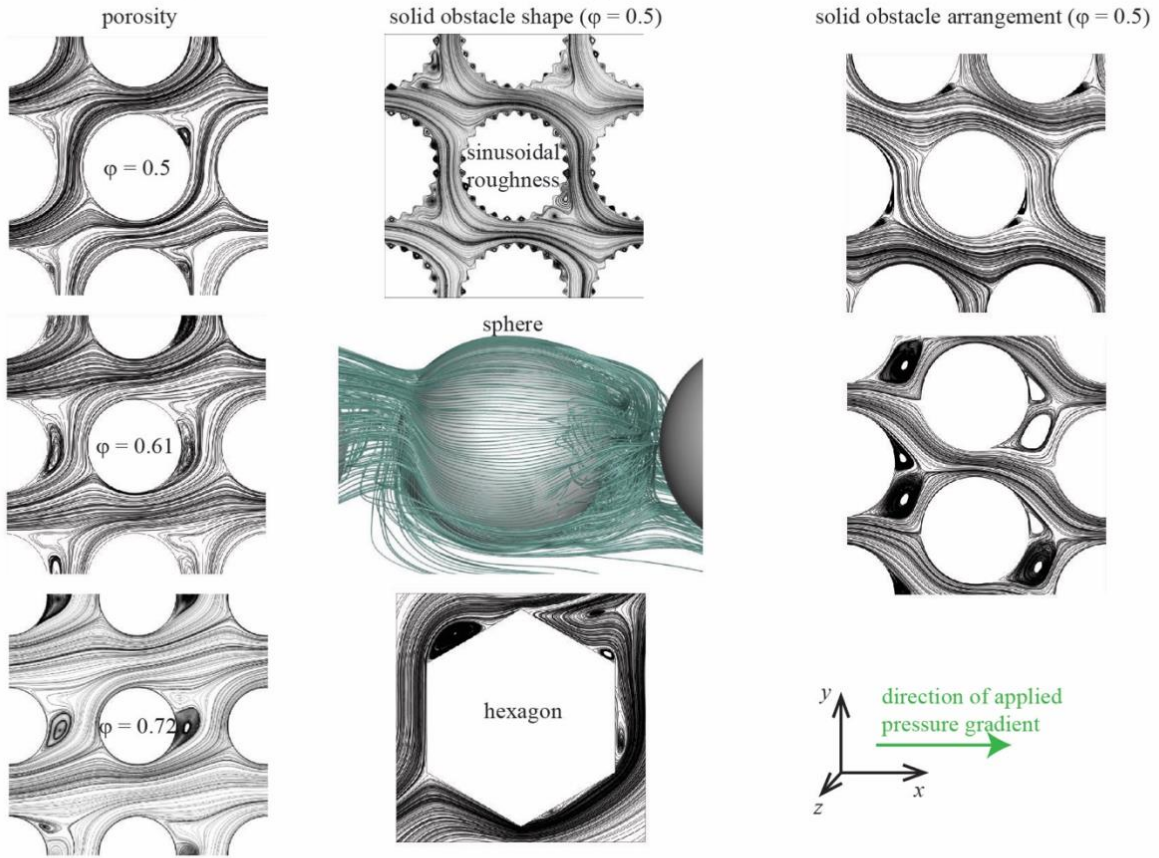

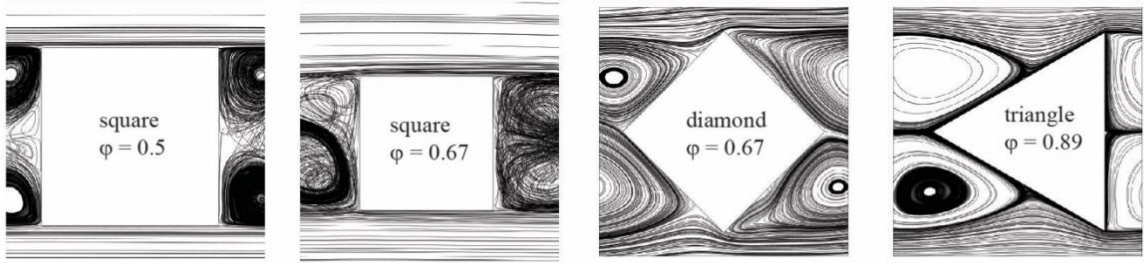

**Fig. 4.** Reynolds averaged flow streamlines for flow at Re = 1000 in porous media with different porosities, solid obstacle shapes, and staggered solid obstacle arrangements, demonstrating which conditions (a) promote symmetry-breaking deviatory flow and (b) which do not.

### 3.2. Symmetry-breaking in the intermediate porosity flow regime (0.8<φ<0.95)

When the porosity of the porous medium is increased to the intermediate porosity regime, there now exists sufficient pore space for dynamic vortex development to occur, but not enough space for the vortices to dissipate in strength before impinging on its downstream neighbor. Asymmetrical flow distribution occurs in this regime that persists in the Reynolds averaged flow distribution, characterized by alternating low and high velocity flow channels forming parallel to the flow direction above and below solid obstacles (Fig. 5). This type of asymmetry was also observed in Kim et al. (2023) in the velocity profile inside the porous layer of a composite porous/fluid channel. Asymmetry in the microscale velocity distribution was caused by biased shedding directions of microscale vortices behind the solid obstacles.

This symmetry-breaking at intermediate porosity only results in a microscale flow deviation, unlike macroscale flow deviation observed in deviatory flow at low porosities, and operates by a completely different mechanism. Unlike at low porosity, the transition from symmetric to asymmetric flow at intermediate porosity occurs between Reynolds numbers of 37 (laminar) and 100 (turbulent), accompanying the transition to turbulence. During laminar flow, a Hopf bifurcation initiates unsteady oscillations, marking the beginning of a secondary flow instability that emerges from interacting shear layers around the obstacle. This Hopf bifurcation is similar to the one observed for square tube arrays for laminar flow (Agnaou et al. 2016). When turbulence develops, the stochastic phase differences in vortex wake oscillations caused by this secondary instability led to flow symmetry breaking. Notably, symmetry breaking does not occur with square cross-section cylindrical obstacles due to their sharp vertices even though the secondary instability is present (Fig. 5). This type of symmetry-breaking is referred to as "jetting flow" adopting the nomenclature used in Ziada (2006) for a similar flow structure.

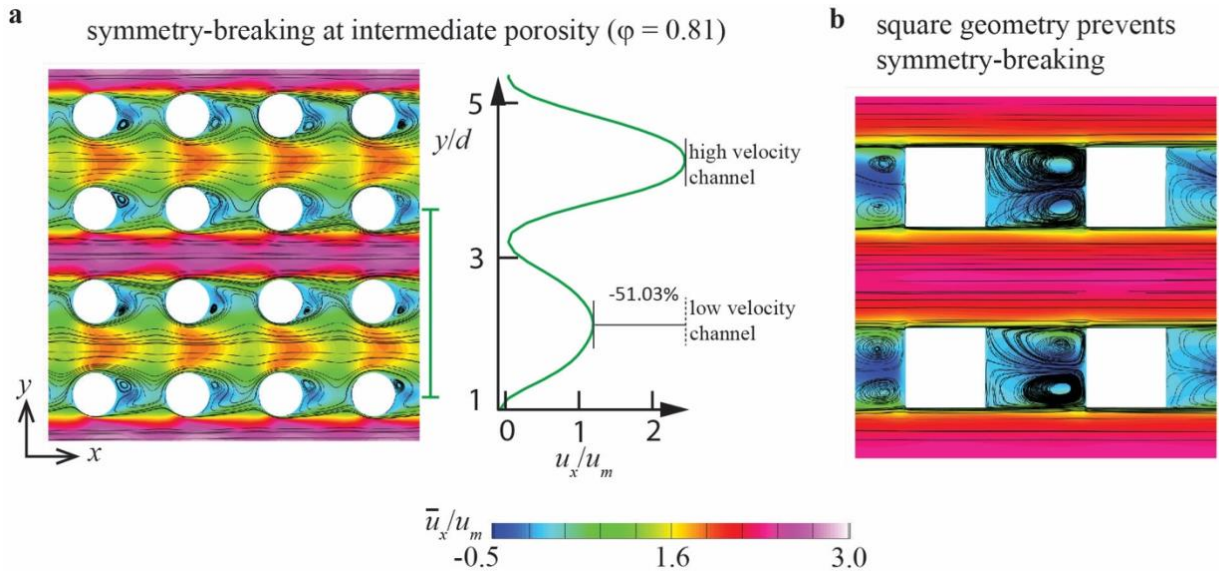

**Fig. 5.** (a) Jetting flow caused by symmetry-breaking in the Reynolds averaged turbulent flow at intermediate porosities at $Re = 300$. Symmetry-breaking causes asymmetric microscale velocity distributions with the formation of high and low velocity channels within the pores. (b) Square cylinder solid obstacles prevent symmetry-breaking from occurring even at higher Reynolds numbers ($Re = 1000$).

The secondary flow instability arises during laminar flow in the porous medium from interactions between shear layers formed around the solid obstacle during microscale vortex formation and downstream solid obstacles. This interaction causes oscillation of the vortex wake behind the obstacle, directing vortex structures into the pore space rather than impinging on neighboring solid obstacle surfaces (Fig. 6(a)). The oscillating vortex wake shifts the separation point on the obstacle surface, reducing vortex recirculation size. Consequently, the flow pattern surrounding the obstacle includes more surface area covered by attached flow.

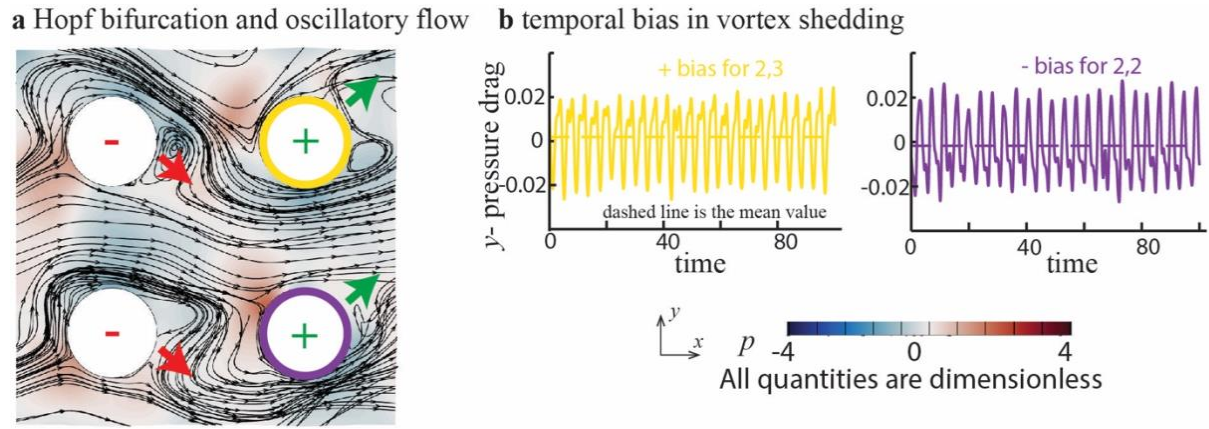

**Fig. 6.** (a) Vortex wake oscillations are caused by the secondary flow instability at φ = 0.82 (*Re* = 300). (b) Wake oscillations are biased at different rows of solid obstacles creating temporal asymmetry in the flow.

Reynolds-averaged velocity and pressure distributions lose symmetry because vortex wake oscillations caused by the secondary flow instability are temporally asymmetric (Fig. 6(b)). When analyzing these oscillations by dividing them into positive and negative phases based on pressure drag force direction acting on the obstacle surface, the positive phase consistently spanned a longer duration than the negative phase (and vice versa depending on the solid obstacle row). This temporal imbalance results in asymmetric Reynolds-averaged velocity and pressure distributions, with velocity distribution showing more prominent asymmetry than the pressure distribution. The velocity channels align with the direction of the applied pressure gradient and alternate in the transverse direction. Low velocity channels form in pore spaces where vortices are preferentially advected due to temporal bias in vortex shedding phases, while high velocity channels develop in complementary pore spaces with less recirculating motion occurring over time. In the pressure distribution, asymmetry appears as shifted stagnation points on the solid obstacle surface away from the symmetric axes for the solid obstacle. For square cylinder solid obstacles, the vertices of the square geometry limit oscillatory vortex wake by preventing separation point shifts, thereby inhibiting symmetry-breaking.

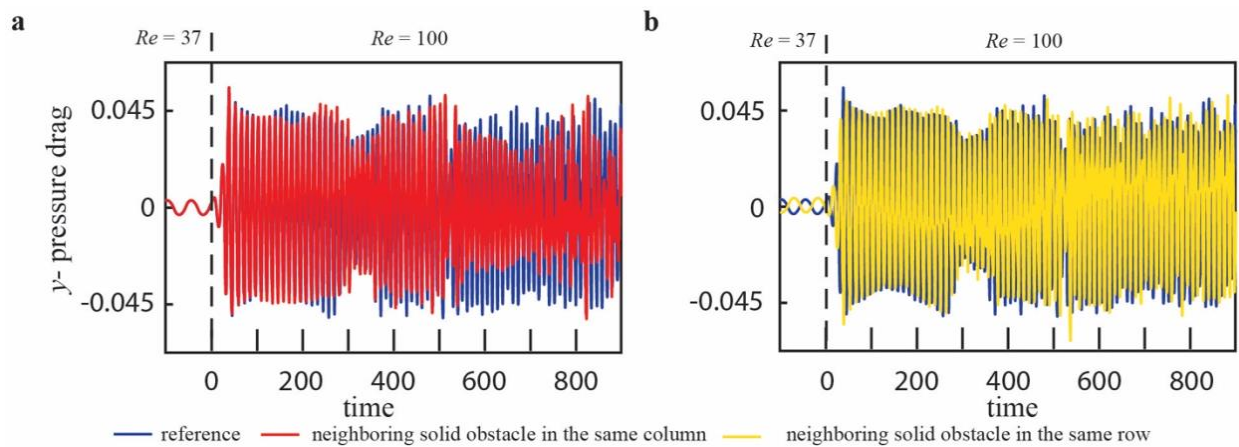

**Fig. 7.** Oscillations of the transverse (*y*-) pressure drag force acting on the individual solid obstacles during the transition from symmetric (*Re* = 37, time < 0) to asymmetric flow (*Re* = 100, time > 0) for porous media with φ = 0.82.

Examination of the flow transition from symmetrical to asymmetrical distribution revealed that symmetry-breaking coincides with transition to turbulence. The secondary flow instability that causes vortex wake oscillation develops in the laminar regime, causing unsteady harmonic oscillations in the flow streamlines around the solid obstacles that is also present in the pressure drag force. Obstacles in the same REV column exhibit zero phase difference in pressure drag force, while obstacles in adjacent columns show exactly one-half period phase difference (Fig. 7). During transition to turbulence in the intermediate porosity flow regime, two significant changes occur: (1) increased amplitudes in spatial flow streamline oscillations and temporal y-pressure drag force oscillations, and (2) emergence of stochastic phase differences in transverse pressure drag force for every obstacle regardless of its REV position. Symmetry-breaking occurs simultaneously with the loss of synchronicity in vortex formation behind REV obstacles, suggesting that symmetry-breaking in turbulent flow at intermediate porosities regulates highly unsteady vortex shedding processes caused by the secondary flow instability. Before symmetry-breaking, vortex wake path oscillation amplitudes are sufficient to interfere with flow around neighboring obstacles in the transverse direction. However, the potentially disruptive combination of high-amplitude vortex wake oscillations and turbulent stochasticity is mitigated by symmetry-breaking. The formation of velocity channels, with vortices preferentially shedding into low velocity channels, regulates the vortex shedding process to prevent vortex collisions.

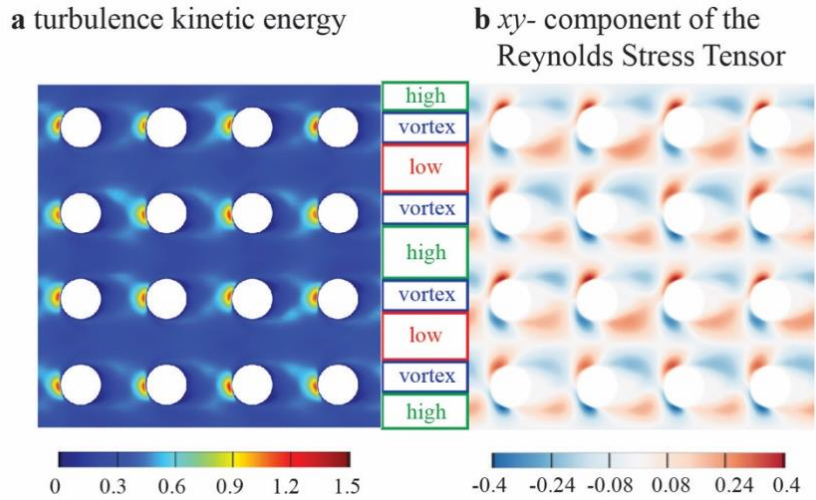

**Fig. 8.** Distributions of the (a) turbulence kinetic energy and (b) $xy$- Reynolds Stress are asymmetrical at the microscale level due to symmetry-breaking ($\varphi = 0.82$, $Re = 300$, circular cylinder solid obstacles).

Asymmetry in the microscale velocity and pressure distributions within the pore space leads to asymmetry in the microscale Reynolds Stress distribution (Fig. 8). Peaks in the turbulence kinetic energy and the $xy$-Reynolds Stress component that are coincident with the high velocity channel have greater magnitude than in the low velocity channel. A shift in the location of the peak turbulence kinetic energy from the plane of geometric symmetry is also noted here. However, this microscale asymmetry in the Reynolds stress is not as significant in magnitude as the microscale $x$- velocity distribution and does not influence the macroscale Reynolds Stress tensor, unlike in the case of deviatory flow at low porosity.

4. **Concluding remarks**

Flow instabilities in porous media exhibit complex behavior and symmetry-breaking bifurcations that vary significantly with porosity, Reynolds number, and solid obstacle geometry. In the low porosity regime ($\varphi <$ 0.8), symmetry-breaking manifests as deviatory flow where the macroscale flow direction deviates from direction of the applied pressure gradient. This phenomenon emerges at higher Reynolds numbers ($Re > 500$) after the transition to turbulence occurs. It results in a bifurcation of the macroscale flow variables like drag force and macroscale Reynolds stress tensor with multiple possible solution states. In contrast, the intermediate porosity regime ($0.8 < \varphi < 0.95$) experiences a different form of symmetry-breaking characterized by alternating high and low velocity channels and temporally biased vortex shedding. This high and low velocity jetting flow phenomenon has a lower critical Reynolds number ($Re=37-100$) coinciding with the flow transition to turbulence.

The foundation for understanding these complex porous media phenomena can be traced back to studies of single and tandem solid obstacles. For isolated obstacles, von Karman vortex shedding begins at $Re \approx 46$, with transition to three-dimensional structures around $Re \approx 190$ and chaotic wakes at $Re > 260$. Proximity to solid walls or other solid obstacles introduces significant modifications to these patterns. For tandem solid obstacles, multiple regimes emerge depending on solid obstacle spacing and arrangement, characterized by dual-frequency dynamics, asymmetric vortex shedding, and phase differences in the vortex motions between neighboring obstacles. These fundamental behaviors of single and tandem obstacles—particularly the mechanisms of vortex interaction, shear layer reattachment, and flow channeling—provide essential insights into the more complex behaviors observed in full porous media, where these mechanisms are amplified and modified by the periodically repeating nature of the solid obstacle arrangement.

Both symmetry-breaking phenomena observed in porous media significantly impact the macroscale properties of turbulent flow in porous media, particularly through the Reynolds stress tensor. At low porosity, the principal axes of the Reynolds Stress Tensor does not align with neither geometric symmetry nor flow direction resulting in anisotropy. However, this anisotropy is balanced by the reduction in the anisotropy of the coherent turbulent structures due to diminished vortex core size. At intermediate porosities, high and low velocity channels with increased and decreased vortical motions, respectively, creates asymmetry in the Reynolds Stress distribution at the microscale level.

The mechanisms underlying these phenomena—competing pressure and inertial forces in low porosity media and temporal bias in vortex shedding in intermediate porosity media—serve important physical functions in regulating flow through confined spaces. These findings highlight the need to consider symmetry-breaking when modeling turbulent flow through porous media, as conventional approaches assuming symmetric flow distributions may not accurately capture the complex dynamics that develop across different porosity regimes. Understanding these mechanisms is crucial for accurately predicting transport phenomena in applications ranging from heat exchangers to urban and environmental flows.

**Statements and Declarations**

**Acknowledgements** The authors acknowledge the computing resources provided by North Carolina State University High Performance Computing Services Core Facility (RRID:SCR_022168). AVK acknowledges the support of the Alexander von Humboldt Foundation through the Humboldt Research Award.

**Funding** This research was funded by the National Science Foundation award CBET-2042834. The National Science Foundation's role in this manuscript is the support of V. Srikanth's doctoral studies.

**Competing Interests** The authors have no relevant financial or non-financial interests to disclose.